\documentclass[preprint,
showpacs,amsmath
,nofootinbib
]{revtex4-1}
\usepackage{graphicx}
\usepackage{dcolumn}
\usepackage{bm}
\usepackage{epstopdf, epsf}
\usepackage{hyperref}
\everymath{\displaystyle}
\begin{document}
\title{Implications of the primordial anisotropy for a scalar field with non-minimal kinetic coupling}
\author{Amir Ghalee}
\email[\,Email:\ ]{ghalee@ut.ac.ir}
\affiliation{{Department of Physics, Tafresh  University,
P. O. Box 39518-79611, Tafresh, Iran}}
\begin{abstract}
We consider a scalar field with a kinetic term non-minimally coupled to gravity in an anisotropic background.
Various potentials for the scalar field are considered. By explicit examples, we show
that how the anisotropy can change the dynamics of the scalar field compared with the isotropic background.
\end{abstract}
\maketitle
\section{\label{sec:level1}INTRODUCTION}
The recent analysis of Planck data showed some deviations from isotropy in the cosmic microwave background \cite{plank-dataa}. It has been argued that
such anomalies are generated by primordial anisotropy components of the space-time metric \cite{lista}. Regarding the last statement, recently study of various cosmological model with a such broken rotational invariant have been considered \cite{lista}.\\
In this work, we study some implications of the primordial anisotropy for a model which is described by the following action
\begin{equation}\label{0-1}
S=\int d^{4}x\sqrt{-g}\left[\frac{R}{2\kappa^{2}}+\frac{1}{2}\lambda^{2}G^{\mu\nu}\partial_{\mu}\varphi\partial_{\nu}\varphi-V(\varphi)
\right].
\end{equation}
Where $\kappa^{2}=8\pi G$ and $G^{\mu\nu}$ is the Einstein's tensor and $\lambda$ is an inverse mass parameter.
Since metric with signature $(-,+,+,+)$ will be used and for the de Sitter space we have $G_{\mu\nu}\propto-g_{\mu\nu}$, we have chosen $+$ sign for the second term. Thus the scalar filed in \eqref{0-1} has the same dynamics as the scalar filed with minimally coupled to gravity in the de Sitter space.\\
The second term in \eqref{0-1} is one of the operators of the Horndeski's scalar-tensor theory \cite{Horndeski}. Thus, in our model we confront with second-order differential equations. Also the second term is the cornerstone of `\textit{The New Higgs Inflation}` \cite{newhiggs}, `\textit{UV-protected Inflation}` \cite{uv} or consider as a curvaton model in Ref. \cite{Feng}. To see consequences of additional terms in \eqref{0-1}, such as the standard kinetic term minimally coupled to gravity, see \cite{list} and references therein.\\
For the anisotropic background metric, following \cite{lista}, we take Bianchi type I metric with the following parametrization
\begin{equation}\label{1-1}
ds^{2}=-N(t)^{2}dt^{2}+e^{2\alpha(t)}\left[e^{-4\sigma(t)}dx^{2}+e^{2\sigma(t)}(dy^{2}+dz^{2})\right],
\end{equation}
where $N(t)$ is the Lapse function, $e^{\alpha}$ and $\sigma$ are the isotropic scale factor and the spatial shear, respectively.\\
By using the ADM formalism, inserting the background metric in the action \eqref{0-1} results in
\begin{equation}\label{1-2}
  s=\frac{1}{2\kappa^{2}}
\int d^{4}xe^{3\alpha(t)}  [\frac{6}{N(t)}(\dot{\sigma}^{2}-\dot{\alpha}^{2})(1-\frac{\lambda^{2}\kappa^{2}\dot{\varphi}^{2}}{2N(t)^{2}})
    -2\kappa^{2}N(t)V(\varphi)].
\end{equation}
By varying the above action with respect to $N$, $\sigma$, $\alpha$ and setting $N=1$, we have

\begin{align}
&(\dot{\alpha}^{2}-\dot{\sigma}^{2})(1-\frac{3}{2}\lambda^{2}\kappa^{2}\dot{\varphi}^{2})= \frac{\kappa^{2}}{3}V(\varphi)\label{1-3},\\ &\ddot{\sigma}(1-\frac{\lambda^{2}\kappa^{2}\dot{\varphi}^{2}}{2})+3\dot{\alpha}\dot{\sigma}(1-\frac{\lambda^{2}\kappa^{2}\dot{\varphi}^{2}}{2})+\dot{\sigma}\frac{d}{dt}(1-\frac{\lambda^{2}\kappa^{2}\dot{\varphi}^{2}}{2})=0\label{1-4},\\
&3(\dot{\sigma}^{2}-\dot{\alpha}^{2})(1-\frac{\lambda^{2}\kappa^{2}\dot{\varphi}^{2}}{2})+6\dot{\alpha}^{2}(1-\frac{\lambda^{2}\kappa^{2}\dot{\varphi}^{2}}{2})\nonumber\\
&+2\frac{d}{dt}\left[\dot{\alpha}(1-\frac{\lambda^{2}\kappa^{2}\dot{\varphi}^{2}}{2})\right]=\kappa^{2}V(\varphi)\label{1-5}.
\end{align}
Also, varying with respect to $\varphi$ yields
\begin{equation}\label{1-6}
\ddot{\varphi}(\dot{\alpha}^{2}-\dot{\sigma}^{2})+3\dot{\alpha}\dot{\varphi}(\dot{\alpha}^{2}-\dot{\sigma}^{2})+\dot{\varphi}\frac{d}{dt}(\dot{\alpha}^{2}-\dot{\sigma}^{2})=-\frac{1}{3\lambda^{2}}\frac{dV(\varphi)}{d\varphi}
\end{equation}
Note that only three of Eqs. \eqref{1-3}-\eqref{1-6} are independent.
\section{\label{sec:level1} Some basic implications}
In Ref. \cite{ghalee} it has been shown that even without potential in the action, \eqref{0-1}, the model give some nontrivial results for the flat Friedmann-Robertson-Walker (FRW) metric. For example one can regard the second term in \eqref{0-1} as an effective description of the dust matter. In this section, we provide reasons to show that the interpretation of the second term is also valid in the anisotropic background. Then we study the consequences of the cosmological constant in the model.
\subsection{\label{sec:level1} pure dust matter}
Recall that in the Einstein's gravity, i.e. $\lambda\rightarrow0$ in \eqref{0-1}, the dust matter is \emph{defined} as a pressureless matter and by definition the pure dust matter is an isotropic matter. Thus, if our interpretation about the second term in \eqref{0-1} is correct, we must obtain the same result with correct expression for the Hubble parameter.\\
Taking $V(\varphi)=0$ in Eq. \eqref{1-4} yields
\begin{equation}\label{2-0}
  \dot{\varphi}^{2}\lambda^{2}\kappa^{2}=\frac{2}{3}
\end{equation}
Therefore, Eqs. \eqref{1-5} and \eqref{1-6} take the following form
\begin{equation}\label{2-1}
\ddot{\sigma}=-3\dot{\alpha}\dot{\sigma},\hspace{.2cm}  \ddot{\alpha}=-\frac{3}{2}(\dot{\alpha}^{2}+\dot{\sigma}^{2}).
\end{equation}
The above equations can be solved as
\begin{equation}\label{2-2}
 \dot{\alpha}=\frac{2}{3t},\hspace{.5cm} \dot{\sigma}=0.
\end{equation}
Thus, by absorbing constants, the metric takes the following form
\begin{equation}\label{2-3}
ds^{2}=-dt^{2}+t^{\frac{4}{3}}(dx^{2}+dy^{2}+dz^{2}).
\end{equation}
So, in this case the Universe behaves as the matter dominated Universe.
The result shows that one can regard the second term in \eqref{0-1} as a pure dust matter.
\subsection{\label{sec:level1} Implications of the cosmological constant}
Observation of the cosmic microwave background and large scale structure are consistent with
an accelerated
expansion phase at present time, which follows after
the dust matter dominated era \cite{plank-data}. The accelerated expansion
phase can be explained by the cosmological constant \cite{plank-data}.\\
To study this case, we take $V(\varphi)=\Lambda$. After some algebra, one can eliminate $\dot{\varphi}$ from Eqs. \eqref{1-3}, \eqref{1-4}, \eqref{1-5}. Then using dimensionless time variable, $\tau\equiv\sqrt{\kappa^{2}\Lambda} t$, the following autonomous equations can be obtained
\begin{align}\label{2-4}
x'=[-54(x^{2}-y^{2})^{3}(x^{2}+y^{2})+9(x^{2}-y^{2})(x^{4}+3y^{4}-4x^{2}y^{2})+3(x^{4}-2y^{4}+x^{2}y^{2})]\big/(x^{2}-y^{2})\Omega\hspace{.1cm},
\end{align}
\begin{equation}\label{2-5}
y'=9xy\left[1-2(x^{2}-y^{2})-12(x^{2}-y^{2})^{2}\right]/\Omega.
\end{equation}
Where the prime denotes derivative with respect to $\tau$ and we have defined the following dimensionless variables
\begin{equation}\label{2-6}
 x\equiv\alpha',\hspace{.2cm} y\equiv\sigma',\hspace{.2cm} \Omega\equiv36(x^{2}-y^{2})^{2}-1.
\end{equation}
The fixed points of the autonomous system are obtained by setting the righthand side of Eqs. \eqref{2-4} and \eqref{2-5} to be $0$. It follows that the autonomous system has two fixed points. One of them is anisotropic fixed point which is determined by
\begin{equation}\label{anf}
x_{aniso}=0,\hspace{.2cm}y_{aniso}=\sqrt{\frac{2}{3}}.
\end{equation}
The stability matrix of the autonomous system around this fixed point is given by
\begin{equation}\label{ans}
\mathcal{M}=\left(
              \begin{array}{cc}
               \frac{\partial x'}{\partial x}& \frac{\partial x'}{\partial y} \\
                \frac{\partial y'}{\partial x} & \frac{\partial y'}{\partial y}\\
              \end{array}
            \right)_{aniso}=\sqrt{6}\left(
                                                                        \begin{array}{cc}
                                                                          0 & \frac{-4}{3} \\
                                                                          -\frac{3}{5} & 0 \\
                                                                        \end{array}
                                                                      \right).
\end{equation}
So, it turns out that the eigenvalues of the stability matrix around the anisotropic fixed point are $+\sqrt{24/5}$ and $-\sqrt{24/5}$.
Since one of the eigenvalues is positive, it turns out that the anisotropic fixed point is not stable. So, the Universe does not
attract to this solution.\\
The other fixed point, which is the isotropic fixed point, is given by
\begin{equation}\label{2-7}
x_{iso}=\frac{1}{\sqrt{3}},\hspace{.2cm}y_{iso}=0.
\end{equation}
The stability matrix  around the isotropic fixed point can be obtained as
\begin{equation}\label{2-8}
\mathcal{M}=\left(
              \begin{array}{cc}
               \frac{\partial x'}{\partial x}& \frac{\partial x'}{\partial y} \\
                \frac{\partial y'}{\partial x} & \frac{\partial y'}{\partial y}\\
              \end{array}
            \right)_{iso}=\sqrt{3}\left(
                                                                        \begin{array}{cc}
                                                                          -2 & 0 \\
                                                                          0 & -1 \\
                                                                        \end{array}
                                                                      \right).
\end{equation}
So, the eigenvalues around of the isotropic fixed point are negative and the fixed point is the attractor fixed point.\\
From the above results it turns out that with any initial condition for the Universe, eventually any anisotropy will be decayed when the
cosmological constant term is presented and finally we have accelerated expansion.
\section{\label{sec:level1} more on the model}
In this section we will investigate dynamics of three type of potentials. For reasons that will soon become clear, we will begin by $V(\varphi)=M^{6}/\varphi^{2}$.
Then we study two type of potentials which are widely use in cosmology, i.e. the quadratic and exponential potential.\\
\emph{Case $V(\varphi)=M^{6}/\varphi^{2}$}: In order to gain some intuition that show how the primordial anisotropic change the dynamics of the model, let us consider $V(\varphi)=M^{6}/\varphi^{2}$.
For the flat FRW background and with this potential , it has been shown that the scalar field is condensed and this phase of the scalar field is the attractor solution \cite{ghalee2}. Also
this phase results in the accelerating expansion for the Universe \cite{ghalee2}. Since it is possible to find an exact solution for this potential and then compare it with the exact solution in the flat FRW background, we study it.\\
Inspiring by the exact solution in the FRW background, let us seek the following solution in the anisotropic background as
\begin{equation}\label{3-0}
\varphi=M^{2}t.\hspace{.25cm} \dot{\alpha}=\frac{1}{\epsilon t},\hspace{.25cm} \dot{\sigma}=\frac{1}{\gamma t}
\end{equation}
Substituting the ansatz into Eq. \eqref{1-4} gives $\epsilon=3$. Note that since one can regard $\dot{\alpha}$ as the Hubble expansion rate in the
anisotropic background, this result indicates that we have decelerating expansion phase. This result is completely different from the isotropic background case. Note that, the results follows from Eq. \eqref{1-4} which is absent in the isotropic background.\\
From Eq. \eqref{1-3} we have
\begin{equation}\label{3-1}
 (\frac{1}{\epsilon^{2}}-\frac{1}{\gamma^{2}})(1-\frac{3}{2}\lambda^{2}\kappa^{2} M^{2})=\frac{\kappa^{2}}{3}M^{2}.
\end{equation}
Finally Eq. \eqref{1-6} is reduced to
\begin{equation}\label{3-2}
 (\frac{1}{\epsilon^{2}}-\frac{1}{\gamma^{2}})=\frac{2}{3\lambda^{2}M^{2}}.
\end{equation}
From Eqs. \eqref{3-1} and \eqref{3-2} we obtain
\begin{equation}\label{3-3}
 \lambda^{2}\kappa^{2} M^{4}=1.
\end{equation}
Eqs. \eqref{3-1}, \eqref{3-3}, with $\epsilon=3$ yield
\begin{equation}\label{3-4}
 \frac{1}{\gamma^{2}}=\frac{2}{3}\kappa^{2}M^{2}+\frac{1}{9}.
\end{equation}
The last result shows the large shear term relative to $\dot{\alpha}$.
But from observation we know that the ratio of $\dot{\sigma}/\dot{\alpha}$ should be small. Thus, the results show that this potential is ruled out when primordial anisotropy is presented.\\
The last but not least note about the stated potential is that, here, the condensed solution is not an attractor solution. To show this point
consider small perturbation $\delta\varphi$ as $\varphi=M^{2}t+\delta\varphi$. Using Eq. \eqref{1-6} we have
\begin{equation}\label{3-5}
\delta\ddot{\varphi}-\frac{\delta\dot{\varphi}}{t}-\frac{3}{t^{2}}\delta\varphi=0.
\end{equation}
The above equation has power low solutions as $\delta\varphi\propto t^{-1}$ and $\delta\varphi\propto t^{3}$. Since the background solution is $\varphi\propto t$, the solution is unstable which is another consequence of primordial anisotropy for the model. Thus, this potential
shows explicitly that how the primordial anisotropy changes the dynamics of the scalar field.\\
To investigate other potentials, note that Eq. \eqref{1-4} can be integrated as
\begin{equation}\label{3-6}
\dot{\sigma}=\frac{1}{(1-\frac{\lambda^{2}\kappa^{2}\dot{\varphi}^{2}}{2})e^{3\alpha}}.
\end{equation}
The above equation shows that, for any potential, the dynamics of the system can be divided into two regimes
\begin{itemize}
  \item $\lambda\kappa\dot{\varphi}\gg1$: For this case as the Universe is expanding, $\dot{\sigma}\approx0$. Thus, this regime is almost similar to isotropy case which studied in other papers \cite{newhiggs,uv,Feng,list,ghalee2}.
  \item $\lambda\kappa\dot{\varphi}\ll1$: From Eq. \eqref{3-6} it follows that $\dot{\sigma}e^{3\alpha}\approx1$. We will focus on this regime.
  For this regime, Eq. \eqref{1-3} takes the following form
\begin{equation}\label{3-7}
  (\dot{\alpha}^{2}-\dot{\sigma}^{2})\approx \frac{\kappa^{2}}{3}V(\varphi).
\end{equation}
Therefore, for this regime, the field equation of motion for $\varphi$ can be written as
\begin{equation}\label{3-8}
  \ddot{\varphi}+3\dot{\alpha}\dot{\varphi}\approx-\frac{V'(\varphi)}{\lambda^{2}\kappa^{2}V(\varphi)}
\end{equation}
\end{itemize}
\emph{Case $V(\varphi)=\frac{1}{2}m^{2}\varphi^{2}$}: For this case, Eqs. \eqref{3-7} and \eqref{3-8} give
\begin{subequations}
\begin{align}
  (\dot{\alpha}^{2}-\dot{\sigma}^{2})\approx &\frac{\kappa^{2}}{6}m^{2}\varphi^{2}\label{3-18},\\
 \ddot{\varphi}+3\dot{\alpha}\dot{\varphi}\approx & \frac{-2}{\lambda^{2}\kappa^{2}\varphi}\label{3-19}.
\end{align}
\end{subequations}
By introducing a new variable $\xi\equiv\varphi^{2}$ and using $\lambda\kappa\dot{\varphi}\ll1$, the above equations take the following form     \begin{subequations}
\begin{align}
  (\dot{\alpha}^{2}-\dot{\sigma}^{2})\approx &\frac{\kappa^{2}}{6}m^{2}\xi\label{3-20},\\
 \ddot{\xi}+3\dot{\alpha}\dot{\xi}\approx & \frac{-2}{\lambda^{2}\kappa^{2}}\label{3-21}.
\end{align}
\end{subequations}
To analyse the above system, firstly we will show that its dynamics is governed by small values of $\xi$. To show this point consider
the opposite side, i.e. $m^{2}\xi\gg1$. So, in Eq. \eqref{3-20} one can neglect $\dot{\sigma}$ compared with $\dot{\sigma}$, that with Eq. \eqref{3-21} results in
\begin{equation}\label{3-22}
\ddot{\xi}+3\frac{\kappa m}{\sqrt{6}}\xi^{\frac{1}{2}}\dot{\xi}\approx\frac{-2}{\lambda^{2}\kappa^{2}}. \hspace{.5cm} ( for \hspace{.2cm} m^{2}\xi\gg1)
\end{equation}
Therefore
\begin{equation}\label{3-23}
 \dot{\xi}\approx\frac{-2}{\lambda^{2}\kappa^{2}}t-2\frac{\kappa m}{\sqrt{6}}\xi^{\frac{3}{2}}. \hspace{.5cm} ( for \hspace{.2cm} m^{2}\xi\gg1)
\end{equation}
Thus, $\dot{\xi}<0$ that shows that even if an initial value for
$\xi$ be a large value, it is driven to a small value.\\
As for $m^{2}\xi\ll1$, Eq. \eqref{3-20} gives $\sigma\approx\alpha$, which shows the large value for primordial anisotropic. This result and $\dot{\sigma}\approx e^{-3\alpha}$ with Eq. \eqref{3-21} leads to
\begin{equation}\label{3-24}
\ddot{\xi}+\frac{\dot{\xi}}{t}\approx\frac{-2}{\lambda^{2}\kappa^{2}}.\hspace{.5cm} ( for \hspace{.2cm} m^{2}\xi\ll1)
\end{equation}
So, $\dot{\xi}=-t/\lambda^{2}\kappa^{2}$. Thus from Eq. \eqref{3-23} it follows that
\begin{equation}\label{3-25}
 \dot{\alpha}=\frac{1}{3t}, \hspace{.35cm} \dot{\sigma}=\frac{1}{t}. \hspace{.5cm} ( for\hspace{.2cm} m^{2}\xi\ll1)
\end{equation}
The results show that the expansion of the universe is decelerating with large anisotropy relative to the Hubble parameter, which is
similar to $V(\varphi)=M^{6}/\varphi^{2}$. \\
\emph{Case $V(\varphi)=V_{0}e^{-\beta\varphi}$}: Substituting this potential into Eqs. \eqref{3-7} and \eqref{3-8} results in
\begin{subequations}
\begin{align}
  (\dot{\alpha}^{2}-\dot{\sigma}^{2})\approx &\frac{\kappa^{2}}{3}V_{0}e^{-\beta\varphi}\label{3-9},\\
 \ddot{\varphi}+3\dot{\alpha}\dot{\varphi}\approx & \frac{\beta}{\lambda^{2}\kappa^{2}}\label{3-10}.
\end{align}
\end{subequations}
Using the following definition
\begin{equation}\label{3-11}
 \frac{\beta^{2}}{4\lambda^{2}\kappa^{2}}\psi^{2}\equiv V_{0}e^{-\beta\varphi},
\end{equation}
Eqs. \eqref{3-9} and \eqref{3-10} take the following form
\begin{subequations}
\begin{align}
(\dot{\alpha}^{2}-\dot{\sigma}^{2})\approx & \frac{\beta^{2}}{4\lambda^{2}\kappa^{2}}\psi^{2}\label{3-12},\\
\ddot{\psi}+3\dot{\alpha}\dot{\psi}\approx&\frac{\beta^{2}}{2\lambda^{2}\kappa^{2}}\psi\label{3-13}.
\end{align}
\end{subequations}
Where $\lambda^{2}\kappa^{2}\dot{\varphi}^{2}\sim\lambda^{2}\kappa^{2}\frac{\dot{\psi}^{2}}{\psi^{2}}$ has been neglected.\\
Eq. \eqref{3-13} can be solved by the WKB approximation as
\begin{equation}\label{3-14}
\psi\approx \exp(\frac{-3}{2}\alpha)(C\cos(\frac{\beta}{\sqrt{2}\lambda\kappa}t)+D\sin(\frac{\beta}{\sqrt{2}\lambda\kappa}t)),
\end{equation}
where $C, D$ are constant.\\
So, from Eqs. \eqref{3-12}, \eqref{3-14} and $\dot{\sigma}e^{3\alpha}\approx1$, it follows that
\begin{equation}\label{3-15}
\dot{\alpha}^{2}\approx e^{-6\alpha}+\frac{\beta^{2}}{4\lambda^{2}\kappa^{2}}e^{-3\alpha}(C\cos(\frac{\beta}{\sqrt{2}\lambda\kappa}t)+D\sin(\frac{\beta}{\sqrt{2}\lambda\kappa}t))^{2}.
\end{equation}
Numerical solution for the above equation is presented in Fig. 1.\\
\begin{figure}
\includegraphics{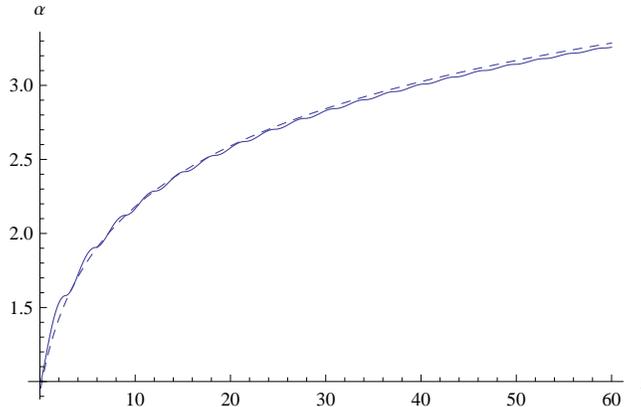}
\caption{\label{fig:epsart}$\alpha$ as a function of time for of the exponential potential. Solid line represents the numerical solution for Eq. \eqref{3-15}. Dashed line shows the result of the analytic expression of Eq. \eqref{3-17}. The parameters are $C=1$, $D=2$, $\beta=\sqrt{2}\lambda\kappa$ and with initial value $\alpha(0.1)=1.$}
\end{figure}
To obtain an analytic expression, we will use the averaging method,
which is widely use in study of nonlinear differential equations \cite{book}. To use the method, note that the righthand side of Eq. \eqref{3-15} is $2\pi\sqrt{2}\lambda\kappa/\beta$-periodic in $t$. In the method of averaging, Eq. \eqref{3-15} is replaced by an averaged expressions when we keep $e^{-\alpha}$ fixed, that results in
\begin{equation}\label{3-16}
\dot{\alpha}^{2}\approx e^{-6\alpha}+(C^{2}+D^{2})\frac{\beta^{2}}{8\lambda^{2}\kappa^{2}}e^{-3\alpha}.
\end{equation}
The above equation can be solved as
\begin{equation}\label{3-17}
e^{3\alpha}=(C^{2}+D^{2})\frac{9\beta^{2}}{32\lambda^{2}\kappa^{2}}(t+t_{0})^{2}-\frac{8\lambda^{2}\kappa^{2}}{(C^{2}+D^{2})\beta^{2}},
\end{equation}
where $t_{0}$ is a constant. As is shown in Fig. 1, the above expression is very good agrement with the numerical solution for Eq. \eqref{3-15}, which indicates the power of averaging method for this case.\\
Thus, the exponential potential results in a universe in which the scale factor, $a\equiv e^{\alpha}$, scales as $a\propto t^{\frac{2}{3}}$, i.e. the dust matter dominated universe but with the shear which behaves as $\dot{\sigma}\propto t^{-2}$.
\section{\label{sec:level1}summary}
In this paper, we have extended the previous works on the model, \eqref{0-1}, to  the anisotropic background.\\
We have shown that, similar to the FRW background, one can regard the second term in \eqref{0-1} as the effective description of the dust matter.
Although with the cosmological constant term we have the anisotropic fixed point as a new feature, but the fixed point is not stable.\\
We have found that how for $V(\varphi)=M^{6}/\varphi^{2}$ the dynamics of the scalar field can be changed dramatically, compared with FRW background. But due to the large value of shear term with respect to the Hubble parameter, this potential can be ruled out.\\
We have demonstrated that the exponential potential results in the matter dominated universe with the small shear term relative to the Hubble parameter. Thus,
if the future data will be agreement with \cite{plank-dataa}, it is reasonable to consider it as toy model for the matter dominated era of the Universe.
\bibliography{apssamp}

\end{document}